\documentstyle[11pt,newpasp,twoside,epsf]{article}
\def\hst{{\it HST}}

\def\edcomment#1{\iffalse\marginpar{\raggedright\sl#1\/}\else\relax\fi}
\def\simgt{\lower 2pt \hbox{$\, \buildrel {\scriptstyle >}\over {\scriptstyle\sim}\,$}}
\def\simlt{\lower 2pt \hbox{$\, \buildrel {\scriptstyle <}\over {\scriptstyle\sim}\,$}}
\reversemarginpar

\begin{document}
\title{\hst\/ Images of Stephan's Quintet: Star Cluster Candidates in a 
Compact Group Environment}
 \author{S. C. Gallagher}
\affil{Dept. of Astronomy \& Astrophysics, Penn State,  
525 Davey Lab, University Park, PA 16802, USA}
\author{S. D. Hunsberger}
\affil{Lowell Observatory, 1400 W. Mars Hill Rd., Flagstaff, AZ 86001, USA}
\author{J. C. Charlton}
\affil{Dept. of Astronomy \& Astrophysics, Penn State}
\author{D. Zaritsky}
\affil{University of Arizona/Steward Observatory, 933 N. Cherry St., Tucson,
AZ 85721, USA}
%
\begin{abstract}
We present \hst\/ WFPC2 images of Stephan's Quintet which encompass three interacting
galaxies and their associated tidal features.  These deep, three-color
($B,V,I$) images indicate recent, massive stellar system formation in
various regions within the compact group environment.  We have identified
star cluster candidates (SCC) both within the interacting galaxies and
in the tidal debris. We compare the SCC colors with stellar population
synthesis models in order to constrain cluster ages, and compare the pattern
of formation of SCC in different regions to the inferred dynamical history  
of the group.
\end{abstract}
%
\section{Introduction}
The Hickson Compact Groups (HCG; Hickson 1982) are among the densest concentrations of
galaxies in the local universe. These high densities combined with relatively low velocity 
dispersions, $\sigma\sim(2-3)\times10^2$~km~s$^{-1}$ (Hickson et al. 1992),
make them active sites of strong galaxy interactions. 
Interactions are believed to initiate bursts of star cluster formation on many
scales from dwarf galaxies along tidal tails
to massive star clusters, the progenitors of today's globular clusters. 
One group in particular, Stephan's Quintet (SQ; also known as HCG~92), is notable for 
evidence of multiple interactions.
\begin{figure}[t]
\plotfiddle{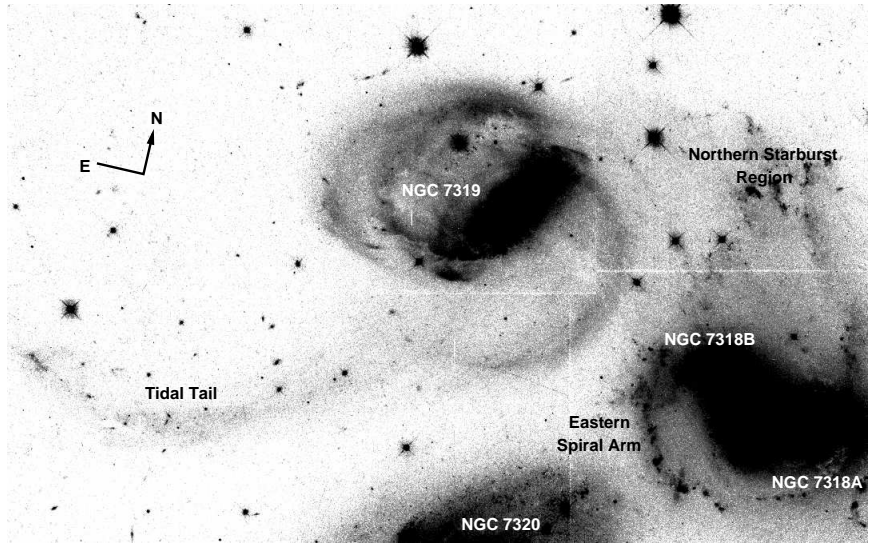}{3in}{-90}{50}{50}{-165}{220}
\caption{This $V-$band image is produced from two overlapping pointings of
WFPC2. The field-of-view is approximately $3\farcm7 \times 2\farcm5$, and the regions of interest
have been labeled.  The fifth member of the group, NGC~7317, is out of the frame to the west.
Note that NGC~7320 is a foreground galaxy not relevant to the current
discussion.}
\end{figure}

SQ is comprised of five galaxies: NGC~7317, NGC~7318A and B, NGC~7319 and NGC~7320
(see Fig.~1 for galaxy identifications).
Based on multiwavelength observations of the group, NGC~7317 and NGC~7320 
show no evidence for recent interactions, unlike the other three galaxies 
(NGC~7320 is a foreground galaxy).
In particular, NGC~7318B shows morphological disruption of spiral structure,
and a long tidal tail extends from NGC~7319.
The interactions have resulted in recent and ongoing star formation
as evident from $B-V$ (Schombert et al. 1990), H$\alpha$
(V\'{\i}lchez \& Iglesias-P\'aramo 1998) and far-infrared (Xu, Sulentic \& Tuffs 1999) 
imaging. 
Furthermore, in the photometric dwarf galaxy study of Hunsberger, 
Charlton, \& Zaritsky (1996), SQ was identified as hosting the richest known system 
of tidal dwarf galaxy candidates.  From these studies, only the largest star-forming 
regions were resolved; many of the young stars appeared to be distributed in the 
diffuse light in the tidal features between the galaxies.  High spatial resolution is 
required to identify star cluster candidates (SCC)
which at the distance of SQ ($z=0.02$; $d\sim66h^{-1}$~Mpc) are faint point sources on the 
Wide Field and Planetary Camera 2 (WFPC2). {\it Hubble Space Telescope} (\hst)
imaging was the obvious next step for investigating the full range in 
scale of massive star formation structure.  Furthermore, with these images 
we could investigate whether 
star clusters form in diverse environments from the inner regions of 
galaxies to tidal debris tens of kiloparsecs from a galaxy center.
%
\section{Observations and Data Analysis}
SQ was observed with the \hst\ WFPC2 in 
two pointings.  The first on 30 Dec 1998, encompassed
NGC~7318A/B and NGC~7319.  The second, on 17 Jun 1999, covered the extended 
tidal tail of NGC~7319.
On both occasions, the images were once dithered{\footnote[1]{Dithering entails 
offsetting the image position
by a half-integer pixel amount in both the $x$ and $y$ directions in order to increase 
the effective resolution of the combined image by better sampling the PSF.  In this case,
we obtained two images in each field and filter.}} 
and taken through three wide-band filters: F450W ($B$), F569W ($V$) and F814W ($I$).    
The exposure times in each field were $4\times 1700$~s, $4\times800$~s and $4\times500$~s for
$B$, $V$ and $I$, respectively.  The data were first processed through the 
standard \hst\ pipeline.  
Subsequently, they were cleaned of cosmic
rays using the STSDAS task GCOMBINE, followed by the IRAF task COSMICRAYS to remove hot
pixels.  Fig.~1 shows the $V$ band image of both fields combined with the
regions of interest labeled.

The initial detection of point sources was undertaken using the DAOFIND routine in 
DAOPHOT (Stetson 1987)
with a very low detection threshold.  This produced thousands of sources per chip, and 
we then performed aperture photometry on all sources.  Those sources with $S/N>3.0$ that 
appeared in the images at both dither positions were retained.  Sources with ${\rm FWHM}>2.5$ 
or $\Delta_V>2.4${\footnote[2]{$\Delta_V$ is the difference between the $V$ magnitudes 
calculated with two photometric apertures: one with radius 0.5~pix and the other 
with radius 3.0~pix.}} 
were rejected as extended (Miller et al. 1997).
Those point sources with $V-I>2.0$ are likely foreground stars, 
and the remaining sources are considered star cluster candidates (SCC).  
This sample will clearly contain some foreground stars and background galaxies, 
but the spatial coincidence of 
most of the sources with the galaxy bulges and tidal features is evidence that
many candidates are legitimate SCC. Approximately 150 sources were found in all 
three filters; they are plotted in the $B-V$ versus $V-I$ color-color plot in Fig.~2.    
In Fig.~3, zoom images
of the tidal tail in NGC~7319 and the northern starburst region (NSR) have the 
SCC marked with circles.  For a discussion of the extended sources in the field, see 
Hunsberger et al. (this proceedings).
\begin{figure}[t!]
\plotfiddle{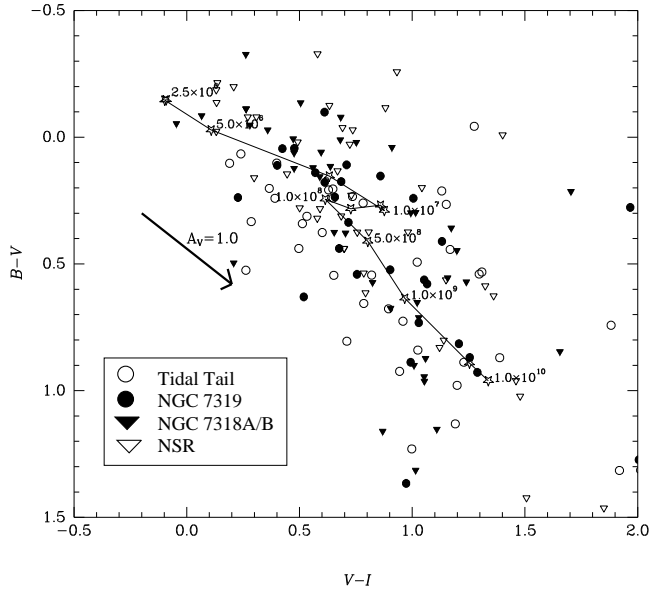}{3.0in}{0}{45}{45}{-130}{-15}
\caption{$B-V$ versus $V-I$ color-color plot of star cluster candidates (SCC).  
The solid line represents the 
evolutionary tracks for a Bruzual \& Charlot (1993) stellar population 
synthesis instantaneous-burst model
(with a Salpeter IMF and solar metallicity).  Numbers along the 
tracks are years.  The SCC photometry has not been corrected for Galactic 
reddening; the models have been reddened with $A_{B}=0.49$
(value from the Large Extragalactic Database for Astronomy; Paturel et al. 1997).}
\end{figure}
\section{Discussion}
\subsection{Dynamical History of Stephan's Quintet}
The diversity of tidal features in SQ is indicative of the complex interaction
history in the group.  In the dynamical history proposed by Moles, Sulentic, \& M\'arquez  
(1997; hereafter MSM97), NGC~7320C (out of the frame of Fig.~1 to the northeast) passed through 
the group a few hundred million years ago stripping NGC~7319 of much of its HI 
(Shostak et al. 1984)
and inducing the extension of the tidal tail.  
In addition, gas was deposited in the area that is currently the NSR.   
This first event would have induced star formation
in the environs of NGC~7319 and perhaps triggered the observed Seyfert~2 
activity in the nucleus.

Two of the four galaxies in Fig.~1, NGC~7319 and NGC~7318A,
have radial velocities within 50~km~s$^{-1}$ of 6600~km~s$^{-1}$.  A third,
NGC~7318B, while apparently interacting with NGC~7318A, has a discordant velocity, 
$v=5700$~km~s$^{-1}$ (Hickson et al. 1992).  This discrepancy is 
inconsistent with the interpretation of NGC~7318B as a foreground galaxy because
of the obvious morphological distortion seen in Fig. 1.
Instead, in the most recent and ongoing interaction event NGC~7318B is 
falling into the group for the first time.
HI maps of the group show that NGC~7318B still retains the bulk of its gas 
(Shostak et al. 1984; MSM97) unlike all of the galaxies with concordant
velocities.  As NGC~7318B approaches SQ, its ISM is shocking the gas of the intragroup medium 
(IGM) in the NSR and along its eastern spiral arm.  An extended arc of both 
radio continuum (van der Hulst \& Rots 1981)
and X-ray (Pietsch et al. 1997) emission supports the 
shocked gas scenario, and H$\alpha$ emission in the same region at the 
radial velocity of NGC~7318B indicates that the collision-compressed gas is being converted
into stars (MSM97).  Hunsberger et al. (1996) also found tidal dwarf 
galaxy candidates along part of the same structure.
\begin{figure}[t!]
\plotfiddle{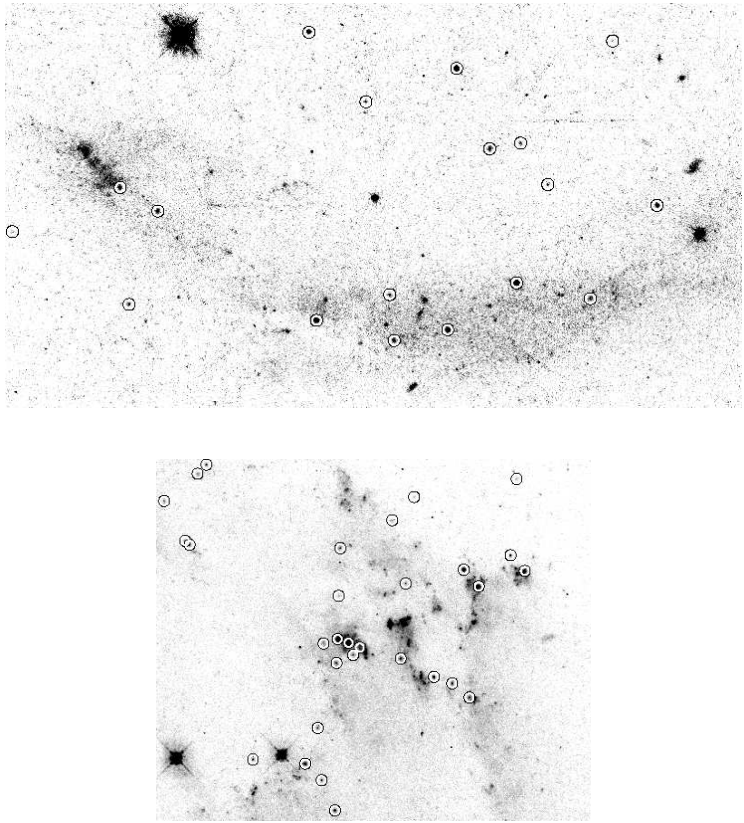}{4.2in}{0}{55}{55}{-150}{0}
\caption{$B-$band zoom images of tidal features with star cluster candidates circled; 
not to the same scale.  {\sc top}: Tidal tail of 
NGC~7319. {\sc bottom}: Northern starburst region.}
\end{figure}
%
\subsection{Star Formation History from SCC Colors}
In all regions with tidal features or galaxies, we identified SCC.
From the simulations of Ashman \& Zepf (1992) of merger remnants, we expected to find
massive young star clusters in the bulges of NGC~7318A and B, but there 
we only detected point sources with colors consistent with old globular clusters.  
This result can be 
understood if the interaction between NGC~7318A and B is relatively recent, and 
star formation is just beginning in the outer regions of the galaxies.  This picture
is consistent with the observations of NGC~7252 (Miller et al. 1997) and 
the Antennae (Whitmore et al. 1999) which suggest that cluster formation is initiated
at large galactic radii and propagates inward over time.  In NGC~7319, we do find young
SCC in the disk and bulge, supporting the older interaction scenario for the event which
stripped it of its gas and pulled out the tidal tail.

From our images, it is also clear that star clusters can form {\it outside} of galaxies. 
In the NSR, the star formation is occurring $\simgt20$~kpc from the bulge of 
the nearest galaxy.  In addition, we discovered several young star clusters in the 
tidal tail of NGC~7319.  In the color-color plot (Fig.~2), there is a clear distinction
between the sources associated with NGC~7318B and those in NGC~7319 and its tidal tail. 
The most recent star formation is occurring in the NSR and the spiral arms of NGC~7318B;
ages of some SCC in those regions are at least as young as 5~Myr.  Any intrinsic dust 
extinction would only cause an overestimate of the ages as the reddening vector is 
approximately parallel to the evolutionary tracks at that point.
In addition to the youngest SCC in each region, we also observe a spread of ages from 
old globular cluster candidates (GCC) with ages $\tau\sim~10^{10}$~yr to more 
intermediate-aged SCC, $\tau\sim~10^{8}$~yr.  This spread is most apparent in the 
NSR and along the tidal tail, both regions where extended periods of 
interaction-induced star formation are reasonable.  Furthermore, the 
presence of the old GCC in the tidal features 
suggests they were pulled out of their birth galaxies as a result of the interactions.  
%
\section{Conclusions}
From \hst\ WFPC2 images, we find $\sim150$ SCC in the environs of SQ.  SCC are 
found both within the bulges of each of the galaxies NGC~7318A/B and NGC~7319,
and also in tidal features. 
The ages deduced from $B-V$ versus $V-I$ colors of SCC
are consistent with the complex interaction scenario outlined by 
MSM97.  Since only old GCC are found in the
centers of NGC~7318A/B, this suggests that recent star formation has not yet occurred there.  Very
young SCC are found along the interaction shock front between the ISM of NGC~7318B 
and the IGM of SQ supporting the hypothesis that this is a recent event.  The spread
of ages in SCC found throughout the field is indicative of recurring episodes of 
interaction-induced star formation.  

\acknowledgements 
We are grateful to A. Kundu and B. Whitmore for sharing 
their expertise in identifying and analyzing point sources in WFPC2 images. 
This work was supported by Space Telescope Science Institute under Grant GO--06596.01.

\section*{Discussion}

{\it J. Gallagher:\/} What is the spatial distribution of the cluster colors as
compared to colors of the more diffuse debris?  This might help in investigating 
differences between cluster formation versus cluster evolution.

\noindent{\it S. G.:\/}  In general, the diffuse emission between the galaxies has $B-V$ colors
similar to those of the outer regions of spiral disks.  More
specifically, in the NSR and along the eastern spiral arm of NGC~7318B, the diffuse light
has $B-V$ colors between 0.3 and 0.5 (Schombert et al. 1990), as do some regions 
in the tidal tail. 
We find young cluster candidates in those regions with similar colors as well 
as some with $B-V<0.3$.

\noindent{\it T. B\"oker:\/} In your color-color diagram, there are a handful of ``clusters'' that 
are not explained by reddening.  Do you have any idea what they are?

\noindent{\it S. G.:\/} There does appear to be a group of point sources clumped below the
evolutionary tracks on the red end of the $V-I$ axis.  I have investigated each of them, and
they do not appear to be part of a distinct population.  A few of these sources
are quite faint in $B$ which could cause some scatter, and there is certainly
some contamination from background galaxies and stars.  

\noindent{\it U. Fritze-von Alvensleben:\/} Where is HI located?  Is there any correlation 
between the absence of HI and the absence of young star clusters?

\noindent{\it S. G.:\/} The HI distribution is unusual as most of the gas in the group 
is {\it outside} of the galaxies.  There is as much HI as is typically found in an 
entire spiral galaxy to the south of NGC~7319, including the tidal tail, 
and a fair amount in the NSB as well
(Shostak et al. 1984).  
We find young SCC in both of those regions. The disk of NGC~7319 is almost entirely 
lacking in gas, but we find some young SCC candidates in that galaxy, though they are
strung along the spiral arms.  The bulge of NGC~7318B still has its HI, and does not
appear to contain any young SCC.

\noindent{\it G. Meurer:\/} Are any of the centers of the galaxies blue?  I suspect the 
reason that you don't see any nuclear clusters is because the galaxies are too far away hence 
crowding makes them difficult to distinguish.

\noindent{\it S. G.:\/}  NGC~7318A and B have similar central colors: $B-V\sim1.0$ and 
$V-I\sim1.2$ that are not particularly blue (though there may be a significant amount
of intrinsic reddening).
NGC~7319 is bluer with $B-V\sim0.5$ and $V-I\sim1.2$; those colors are consistent with
the Seyfert~2 activity in the nucleus. The complex structure in the center of 
each of these galaxies would certainly make detecting a nuclear cluster very difficult. 
However, we find no young SCC within the inner 2--3~kpc even where the light distribution 
is smooth.  In NGC~7319 we do find young SCC within the bulge of the galaxy.

\end{document}